\documentstyle[sprocl]{article}

\input{epsf} 

\def\PRL{{\em Phys.~Rev.~Lett.}}
\def\PRD{{\em Phys.~Rev.}~D}
\def\EPJC{{\em Eur.~Phys.~J.}~C}

\def\ra{\rightarrow}
\def\be{\begin{equation}}
\def\ee{\end{equation}}
\def\bea{\begin{eqnarray}}
\def\eea{\end{eqnarray}}

\newcommand{ \centeron }[2]{{\setbox0=\hbox{#1}\setbox1=\hbox{#2}\ifdim
                             \wd1>\wd0\kern.5\wd1\kern-.5\wd0\fi \copy0
                             \kern-.5\wd0\kern-.5\wd1\copy1\ifdim\wd0>\wd1
                             \kern.5\wd0\kern-.5\wd1\fi}}
\newcommand{ \ltap }{\;\centeron{\raise.35ex\hbox{$<$}}
                     {\lower.65ex\hbox{$\sim$}}\;}
\newcommand\G{ \tilde{G} }

\newcommand{\sss}{\scriptscriptstyle}
\newcommand{ \tauu       }{ { \tilde \tau}_1 }
\newcommand{ \NI         }{ \tilde{N}_1 }
\newcommand{\epem}{e^{\sss +}e^{\sss -}}   

\begin{document}

\begin{titlepage} 
\thispagestyle{empty} 

\begin{flushright} 
hep-ph/9907396 \hfill CERN-TH/99-214 \\
\end{flushright}

\vspace{0.2cm}

\hrule\hfill

\vspace{-0.1cm} 

\hrule\hfill

~\\ 

\begin{center}
{\Large \bf Precision GMSB at a Linear Collider$^{\: \star}$} \\

~\\

{\large Sandro~Ambrosanio} \\

CERN -- {\it Theory Division}, \\
CH-1211 Geneva 23, \\ 
Switzerland \\
e-mail: {\sl ambros@mail.cern.ch} \\

~\\ 

{\large and} 

~\\ 

{\large Grahame~A.~Blair} \\

{\it Royal Holloway and Bedford New College}, \\
      University of London, Egham Hill, Egham, \\
      Surrey TW20 0EX, United Kingdom \\
      e-mail: {\sl g.blair@rhbnc.ac.uk} \\

\vspace*{\fill} 
{\bf Abstract} \\ 
\end{center} 
{\small 
We simulate precision measurements of gauge-mediated supersymmetry breaking 
(GMSB) parameters at a 500 GeV $\epem$ linear collider in the scenario where 
a neutralino is the next-to-lightest supersymmetric particle. 
Information on the supersymmetry breaking and the messenger sectors of the 
theory is extracted from the measured sparticle mass spectrum and 
neutralino lifetime. 
}

\vspace*{\fill} 

\begin{center} 
Worldwide Study on Physics and Experiments with Future Linear $\epem$ 
Colliders \\
4$^{\rm th}$ International Workshop on Linear Colliders (LCWS'99) \\
Sitges, Barcelona, Spain, April 28 -- May 5, 1999. 
\end{center}

\vspace*{\fill} 

\noindent 
\parbox{0.4\textwidth}{\hrule\hfill} \\ 
{\small 
$^\star$\ Work supported also by {\it Deutsches Elektronen-Synchrotron} DESY, 
          Hamburg, Germany.}  

\end{titlepage}

\setcounter{page}{0}

\thispagestyle{empty} 
~\\
\newpage
\thispagestyle{plain}

\title{ PRECISION GMSB AT A LINEAR COLLIDER~\footnote{Work 
supported also by DESY, Hamburg.} } 

\author{ S.~AMBROSANIO }

\address{ CERN, Theory Division, CH-1211 Geneva 23, Switzerland }

\author{ G.~A.~BLAIR }

\address{ Dept. of Physics, RHBNC, Egham, Surrey TW20 0EX, U.K. } 

\maketitle\abstracts{We simulate precision measurements of 
gauge-mediated supersymmetry (SUSY) breaking (GMSB) parameters at 
a 500 GeV $\epem$ linear collider (LC) in the scenario where a 
neutralino is the next-to-lightest supersymmetric particle (NLSP). 
Information on the SUSY breaking and the messenger sectors of the 
theory is extracted from the measured sparticle mass spectrum and 
neutralino lifetime. 
}
  
\noindent
GMSB~\cite{GRreport} is an attractive possibility for physics beyond 
the standard model (SM) and provides natural suppression of the SUSY 
contributions to flavour-changing neutral currents at low energies. 
In GMSB models, the gravitino $\G$ is the LSP with mass given by 
$m_{\G} = \frac{F}{\sqrt{3}M'_P} \simeq 
2.37 \left(\frac{\sqrt{F}}{100 \; {\rm TeV}}\right)^2 \; {\rm eV}$,  
where $\sqrt{F}$ is the fundamental SUSY breaking scale. 
The GMSB phenomenology is characterised by decays of the NLSP to its 
SM partner and the $\G$ with a non-negligible or even macroscopic 
lifetime. 
In the simplest GMSB realizations, depending on the parameters
$M_{\rm mess}$, $N_{\rm mess}$, $\Lambda$, $\tan\beta$, sign($\mu$)
defining the model, the NLSP can be either the lightest neutralino 
$\NI$ or the light stau $\tauu$. For this study~\cite{ourpaper},
we generated several thousand GMSB models following the standard 
phenomenological approach~\cite{AKM} and focused on the neutralino
NLSP scenario, for which we selected several representative points
for simulation. Our aim was to explore the potential of a LC
in determining the GMSB parameters. Firstly, for a sample model with light 
sparticles, we considered a measurement of the spectrum via threshold 
scanning and found that $N_{\rm mess}$ and $\Lambda$ ($M_{\rm mess}$ and 
$\tan\beta$) can be determined to 0.1 (1)\% after a 200 fb$^{-1}$ run with 
c.o.m. energy between 200 and 500 GeV. Then, we investigated $\NI$ lifetime 
measurements in the whole allowed $c\tau_{\NI}$ range, performing detailed 
event simulation for a set of representative GMSB models. 
Indeed, since the $\NI$ lifetime is related to $\sqrt{F}$ by
\be
c \tau_{\NI} = \frac{16\pi}{{\cal B}} \frac{\sqrt{F}^4}{m_{\NI}^5}
\simeq \frac{1}{100 {\cal B}} 
\left(\frac{\sqrt{F}}{100 \; {\rm TeV}}\right)^4 
\left(\frac{m_{\NI}} {100 \; {\rm GeV}}\right)^{-5},
\label{eq:NLSPtau}
\ee 
the GMSB framework provides an opportunity to extract information on the 
SUSY breaking sector of the theory from collider experiments that is not 
available, e.g., in supergravity-inspired models. 

\begin{figure}
\centerline{
\epsfxsize=3.0in
\epsffile{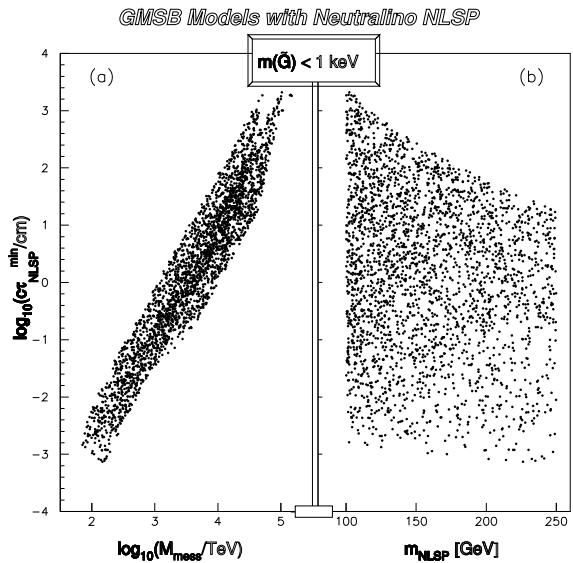} 
}
\caption{Scatter plot of the neutralino NLSP lifetime as a function of 
the messenger scale $M_{\rm mess}$ (a) and $m_{\NI}$ (b). 
For each set of GMSB model input parameters, we plot the lower limit on 
$c\tau_{\NI}$, corresponding to 
$\sqrt{F} \simeq \sqrt{F}_{\rm mess} = \sqrt{\Lambda M_{\rm mess}}$. 
We use only models that fulfil the limit 
$m_{\G} \ltap 1 \; {\rm keV} \Rightarrow 
\sqrt{F}_{\rm mess} \ltap \sqrt{F} \ltap 2000$ TeV 
suggested by simple cosmology. 
}
\label{fig:one}
\end{figure}

 For the models we considered, the neutralino lifetime ranges from microns
to tens of metres. While the lower bound on $c\tau_{\rm \NI}$ comes from 
requiring perturbativity up to the grand unification scale~\cite{AKM}, the 
upper bound is only valid if the $\G$ mass is restricted to be lighter than 
about 1 keV, as suggested by some cosmological arguments \cite{Cosmo}. 
This circumstance is summarised in Fig.~\ref{fig:one}.

\begin{figure}
\centerline{
\epsfxsize=3.0in
\epsffile{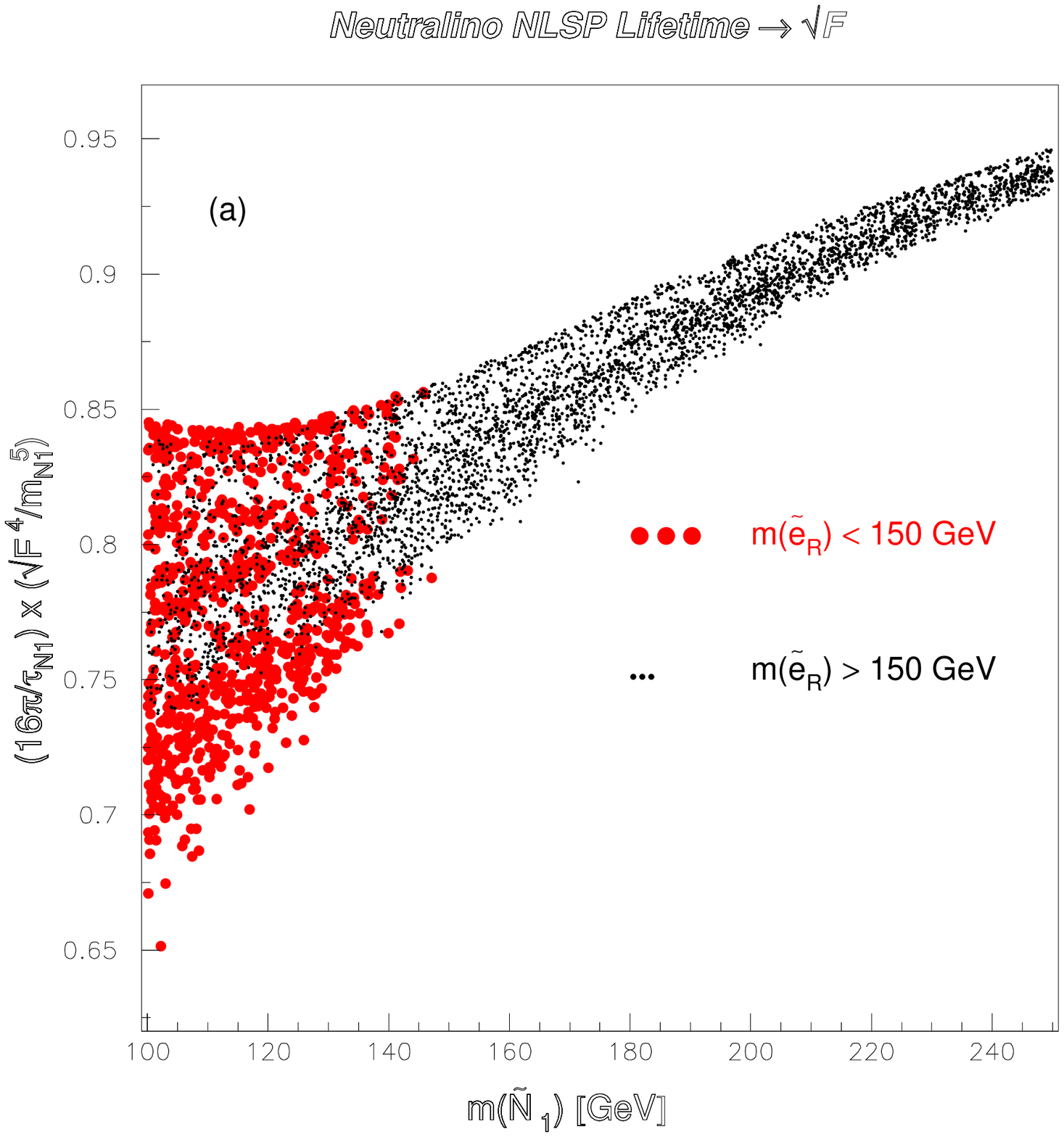} 
\hfill
\epsfxsize=3.0in
\epsffile{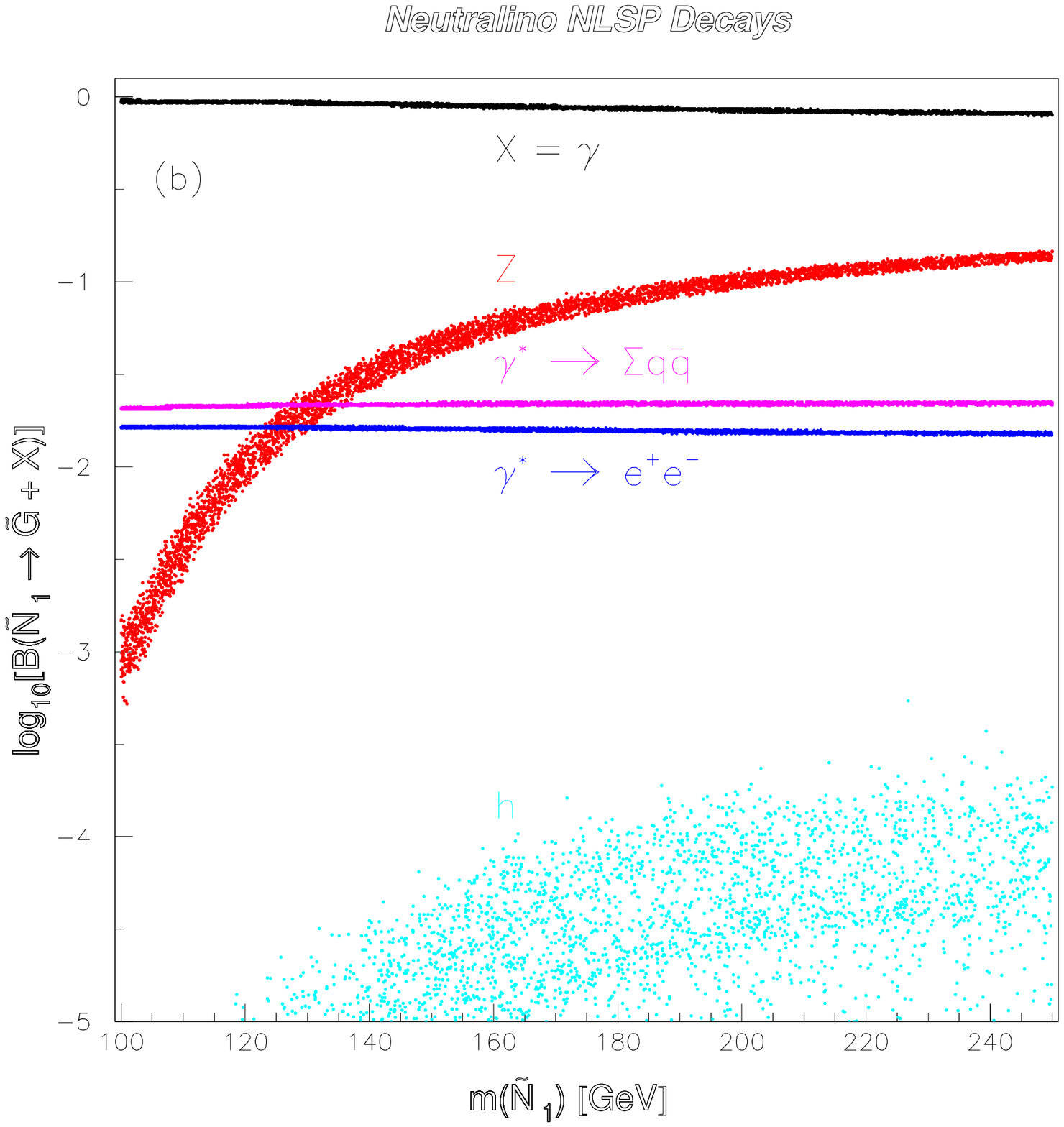} 
}
\caption{(a) Scatter plot showing the relation between the neutralino NLSP
lifetime and the fundamental scale of SUSY breaking $\sqrt{F}$, i.e. the 
factor ${\cal B}$ in Eq.(\ref{eq:NLSPtau}), as a function of the 
neutralino mass in GMSB models of interest for the LC 
(100 $\ltap m_{\NI} \ltap 250$ GeV).  
Big grey dots in represent models with a light $R$-selectron (102--150 GeV), 
small black dots are for the heavier selectron 
case (150--430 GeV). 
(b) Scatter plot for the BR's of various $\NI$ decay channels as 
a function of the $\NI$ mass. Dots in different grey scale (colours) refer 
to the decays $\NI \ra \gamma\G$, $\NI\ra Z \G$ (including off-shell effects), 
and to hadrons or $\epem$ plus gravitino via virtual photon, as labelled. 
For reference, we also report results for the 2-body $\NI\to h^0\G$ decay 
in the on-shell approximation, whose BR is always negligible.} 
\label{fig:two}
\end{figure}

 For given $\NI$ mass and lifetime, the residual theoretical uncertainty 
on determining $\sqrt{F}$ is due to the factor of order unity ${\cal B}$ 
in Eq.~(\ref{eq:NLSPtau}), whose variation is quite limited in GMSB 
models (cfr. Fig.~\ref{fig:two}a). 

 For our simulations, and in particular for short $c\tau_{\NI}$ 
measurements, it was fundamental to take the $\NI\ra\G f \bar{f}$ decays  
into account, in addition to the dominant one $\NI\ra\gamma\G$. 
We performed a complete analysis of these channels and found that in most
cases of interest for our study the total width is given approximately by 
\be
\Gamma(\NI\ra f\bar{f}\G) = \Gamma(\NI\ra\gamma\G) 
\frac{\alpha_{\rm em}}{3\pi} N_f^c Q_f^2 
\left[2 \; {\rm ln}\frac{m_{\NI}}{m_f} - \frac{15}{4} \right]
+ \Gamma(\NI\ra Z\G)B(Z\ra f\bar{f}) \; , 
\label{eq:Steve-for}
\ee
where the expressions for the widths of the 2-body $\NI$ decays 
are well-known~\cite{AKKMM}. In Fig.~\ref{fig:two}b, the branching ratio
(BR) of the $\NI\ra \gamma\G$ decay is compared to those of $\NI\ra Z\G$ 
and $\NI\ra h^0 \G$ (in the on-shell approximation) and those of the main 
$\NI\ra f \bar{f}\G$ channels (including virtual-photon exchange 
contributions only). 

 Using {\tt CompHEP 3.3.18}~\cite{CompHEP} together with a home-made 
lagrangian including the relevant gravitino interaction vertices in 
a suitable approximation, we also studied the kinematical distributions 
of the $\NI\to f \bar{f}\G$ channels and implemented the numerical results 
in our GMSB event generator. 

 For this purpose, we used {\tt SUSYGEN 2.2/03}~\cite{susygen}, modified to 
include the 3-body neutralino decays as discussed above. 
For each sample GMSB model, we considered in most cases a LC run at a 
c.o.m. energy such that the only SUSY production process open is 
NLSP pair production $\epem\ra\NI\NI$, followed by $\NI$ decays through  
all possible channels. For more challenging models where the light SUSY 
thresholds are close to each other, we simulated also events from
$R$-slepton pair production and used some selection cuts to isolate the 
$\NI\NI$ events, for which the $\NI$ production energy is fixed by the 
beam energy (we also took into account initial-state radiation as well
as beamstrahlung effects), allowing a cleaner $c\tau_{\NI}$ measurement. 

 The primary vertex of the events was first smeared according to the assumed 
beamspot size of 5 nm in $y$, 500 nm in $x$ and  400 $\mu$m in $z$ and then 
the events were passed through a full {\tt GEANT 3.21}~\cite{Geant321} 
simulation of the detector as described in the ECFA/DESY CDR~\cite{CDR}.
The tracking detector components essential to our analysis included a 
5-layer vertex detector with a point precision of 3.5 $\mu$m in $r\phi$ 
and $z$, a TPC possessing 118 padrows with point resolution of 160 $\mu$m 
in $r\phi$ and 0.1 cm in $z$.
In addition we assumed an electromagnetic calorimeter with energy resolution
given by ($10.3/\sqrt{E} + 0.6$)\%, angular pointing resolution of
$50/\sqrt{E}$ mrad and timing resolution of $2/\sqrt{E} $ ns.  
The dimensions of the whole calorimeter (electromagnetic and hadronic) 
were 172 cm $< r < 210$ cm and 280 cm $ <|z| < 330$ cm.  

 A single neutralino produced with energy $E_{\NI}$ will decay before
travelling a distance $\lambda$ with a probability given by 
$P(\lambda) = 1 - {\rm exp}(-\lambda/L)$, where $L = c \tau_{\NI} 
(\beta\gamma)_{\NI}$ is the  $\NI$ ``average'' decay length and 
$(\beta\gamma)_{\NI} = (E_{\NI}^2/m_{\NI}^2 - 1)^{1/2}$.

\begin{figure}
\centerline{
\epsfxsize=3.0in
\epsffile{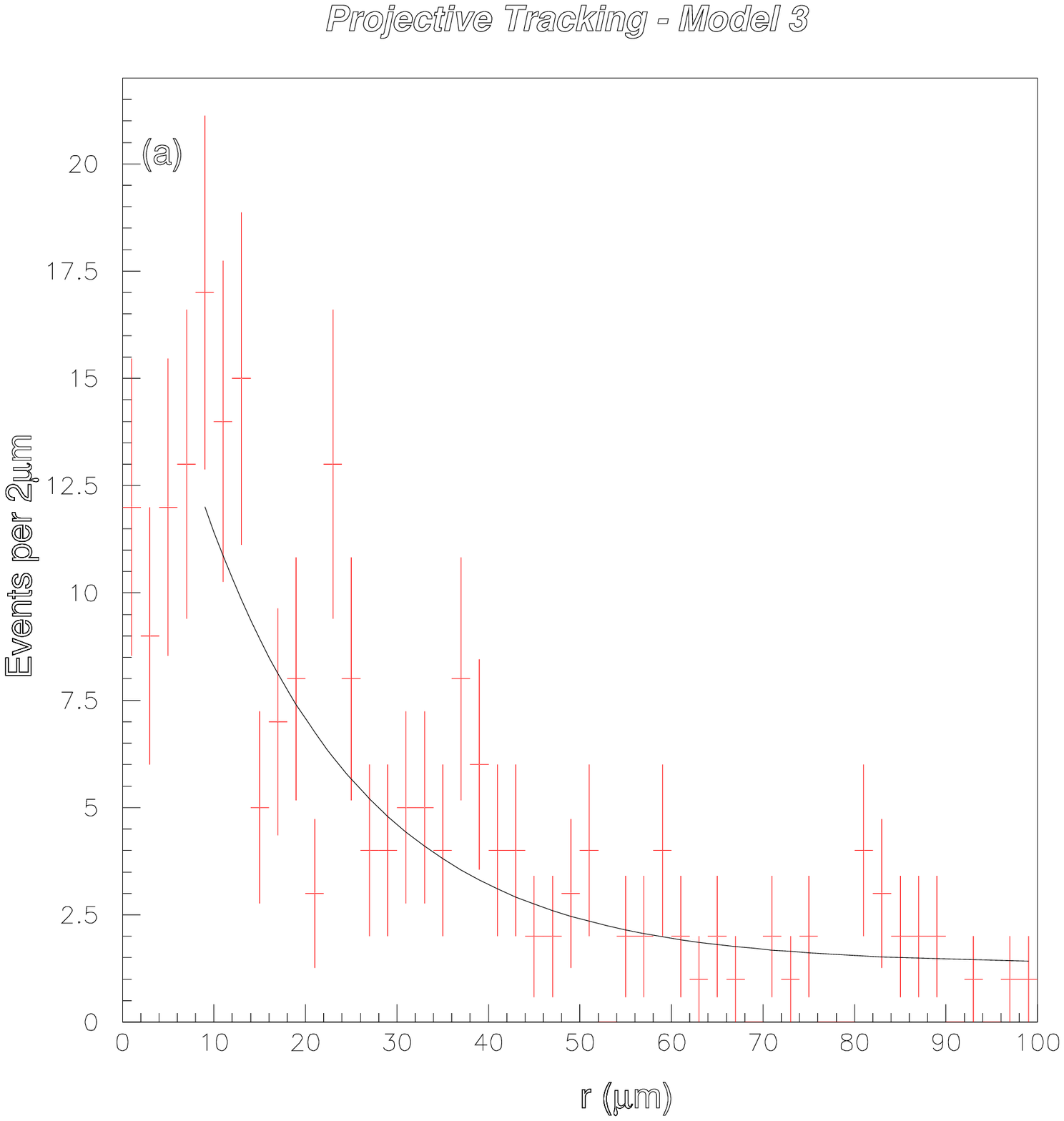} 
\hfill
\epsfxsize=3.0in
\epsffile{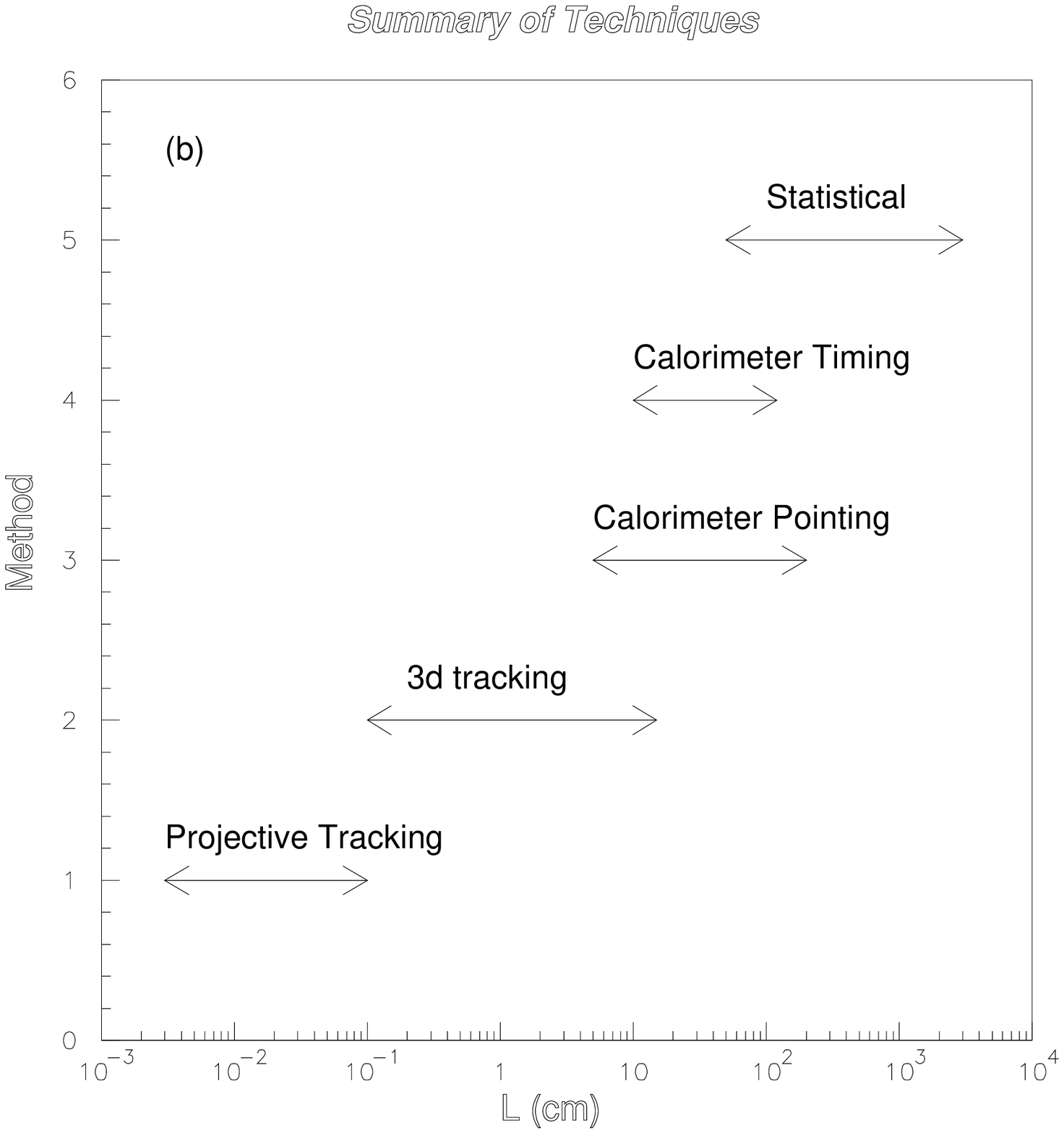} 
}
\caption{(a) Reconstructed projective radial distances, $r$, of the 
$\NI\ra f \bar{f}\G$ decay vertex, for a LC run on a short-lifetime model  
with $L = 10$ $\mu$m together with a fit to an exponential plus constant. 
(b) Summary of the techniques used here for a $c\tau_{\NI}$ measurement 
to 10\% or better.} 
\label{fig:three}
\end{figure} 

 For $L$ less than a few cm, we used tracking for measuring 
the vertex of $\NI\ra\G f \bar{f}$ decays. 
For the very short case, $L$ less than a few hundred $\mu$m, the beamspot
size becomes important and a 3D procedure is not appropriate.  
Instead, the reconstructed vertex was projected onto the 
$xy$ plane, where the beamspot size is very small, and we 
used the resulting distributions to measure the $\NI$ lifetime.  
We studied several GMSB models with $c\tau_{\NI}$ in the allowed range
and found that the intrinsic resolution of the method was approximately 
10 $\mu$m. An example of the reconstructed 2D decay length distribution 
for a challenging model where the neutralino lifetime can be very 
short~\cite{ourpaper} is shown in Fig.~\ref{fig:three}a for statistics 
corresponding to 200 fb$^{-1}$ ($r$ is the $xy$ component of $\lambda$).  

 For 500 $\mu$m $\ltap L \ltap 15$ cm, we used 3D vertexing to determine 
the decay length distribution and hence the lifetime of the $\NI$.  
Vertices arising from $\NI\ra \gamma\G$ and photon conversions in detector 
material were essentially eliminated using cuts on the invariant mass of 
the daughter pairs together with geometrical projection cuts involving the 
mass of the $\NI$ and the topology of the daughter tracks. Methods of 
measuring the $\NI$ mass using the endpoints of photon energies or
threshold techniques, together with details of the projection cuts
have been described~\cite{ourpaper}. Using 200 fb$^{-1}$ of data, 
we concluded that a lifetime measurement with statistical error
of approximately 4\% could be made using this method.

 For $L$ larger than a few cm, we used the $\NI\ra\gamma\G$ channel.  
The calorimeter was assumed to have pointing capability, using the
shower shapes together with appropriate use of pre-shower detectors.
Assuming the pointing angular resolution mentioned above, we 
demonstrated~\cite{ourpaper} how a decay length measurement can be made. 
We concluded that for lifetimes ranging from approximately 5 cm to 
approximately 2 m this method worked excellently, with statistical precisions 
ranging from a few \% at the shorter end to about 6\% at the upper end of the 
range.
We also investigated the use of timing information to provide a lifetime
measurement but found it to be of less use than calorimeter pointing.  
However, the use of timing in assigning purely photonic events to bunch 
crossings and for rejecting cosmic backgrounds should not be 
underestimated.

 For very long lifetimes, we employed a statistical technique where
the ratio of the number of one photon events in the detector to the 
number of two photon events was determined as a function of $c\tau_{\NI}$, 
requiring the presence of at least one displaced $\gamma$.   
This allowed a largely model-independent measurement out to 
$c\tau_{\NI} \simeq$ few 10's m. The possibility of using the ratio of the 
number of no-photon SUSY events to one photon events was also 
discussed~\cite{ourpaper}. The latter allows a greater length reach, but 
relies on model-dependent assumptions.

 In Fig.~\ref{fig:three}b, we summarise the techniques we have used as 
a function of $L$ for a sample model. The criterion for indicating a 
method as successful is a measurement of $L$ and the $\NI$ lifetime to 
10\% or better. It can be seen that $L$ can be well measured for  
10's of $\mu$m $\ltap L \ltap$ 10's of m, which is in most cases enough 
to cover the wide range allowed by theory and suggested by cosmology. 

 With reference to Eq.~(\ref{eq:NLSPtau}), we note that a 10\% error
in $c\tau_{\NI}$ corresponds to a 3\% error in $\sqrt{F}$.
This is of the same order of magnitude as the uncertainty on the factor 
${\cal B}$, which parameterises mainly the different possible $\NI$ physical 
compositions in GMSB models (cfr. Fig.~\ref{fig:two}a). 
We also checked explicitly that, in comparison, the contributing error from 
a neutralino mass measurement using threshold-scanning techniques or 
end-point methods is negligible~\cite{ourpaper}. 

 Hence we conclude that, for the models considered and under 
conservative assumptions, a determination of $\sqrt{F}$ with a precision 
of approximately 5\% is achievable at a LC simply by performing $\NI$ 
lifetime and mass measurements in the context of GMSB with neutralino NLSP. 
Less model dependent and more precise results can be obtained by 
adding information on the $\NI$ physical composition from other observables, 
such as $\NI$ decay BR's, cross sections, distributions.

\end{document}